\documentclass{aa}  

\usepackage{ulem}

\usepackage{graphicx}
\usepackage{txfonts}
\begin{document}

   \title{Mass-loading of the solar wind at 67P/Churyumov-Gerasimenko}
   \subtitle{Observations and modelling}

   \author{E. Behar
          \inst{1,2},
          J. Lindkvist
          \inst{1,3},
          H. Nilsson
          \inst{1,2},
          M. Holmstr\"{o}m
          \inst{1},
          G. Stenberg-Wieser
          \inst{1},
          R. Ramstad
          \inst{1,3},
          C. G\"{o}tz
          \inst{4}}
   \authorrunning{E. Behar et al.}
          
   \institute{Swedish Institute of Space Physics, Kiruna, Sweden.\\
               Lule\aa\ University of Technology, Department of Computer Science, Electrical and Space Engineering, Kiruna, Sweden\\
               Ume\aa\ University, Department of Physics, Ume\aa, Sweden\\
               Technicsche Universit\"{a}t Braunschweig, Institute for Geophysics and Extraterrestrial Physics, Braunschweig, Germany.\\
              \email{etienne.behar@irf.se}             }

   \date{Received 26 April 2016; accepted 12 September 2016}

% \abstract{}{}{}{}{} 
% 5 {} token are mandatory
 
  \abstract
   {The first long-term in-situ observation of the plasma environment in the vicinity of a comet, as provided by the European Rosetta spacecraft.}
   {Here we offer characterisation of the solar wind flow near 67P/Churyumov-Gerasimenko (67P/CG) and its long term evolution during low nucleus activity.  We also aim to quantify and interpret the deflection and deceleration of the flow expected from ionization of neutral cometary particles within the undisturbed solar wind. }
   {We have analysed {\it in situ} ion and magnetic field data and combined this with hybrid modeling of the interaction between the solar wind and the comet atmosphere.}
   {The solar wind deflection is increasing with decreasing heliocentric distances, and exhibits very little deceleration. This is seen both in observations and in modeled solar wind protons. According to our model, energy and momentum are transferred from the solar wind to the coma in a single region, centered on the nucleus, with a size in the order of 1000 km. This interaction affects, over larger scales, the downstream modeled solar wind flow. The energy gained by the cometary ions is a small fraction of the energy available in the solar wind.}
   {The deflection of the solar wind is the strongest and clearest signature of the mass-loading for a small, low-activity comet, whereas there is little deceleration of the solar wind.}


   \maketitle
   
\section{Introduction}

 Comets show large changes in their appearance as their distance from the Sun varies. Closer to the Sun, volatile materials on the comet nucleus start to sublimate, forming a neutral cloud that becomes partially ionized by solar UV radiation and charge exchange processes. 
When the comet activity evolves, the complexity of its interaction with the solar wind also changes. At large distances from the Sun the solar wind directly impacts the surface of an atmosphereless nucleus in an asteroid-like interaction, while at smaller heliospheric distances the solar wind permeates a thin, partially ionized, unstructured coma. When the comet activity is even higher (or the comet is closer to the Sun), the coma is much denser and plasma boundaries form, creating a cometary magnetosphere (\citet{szego2000ssr} section 4.1, \citet{koenders2015pss}). 
  
 {\it In situ} investigations of the interaction between active comets and the solar wind started in the mid-80s with the International Cometary Explorer (ICE) visiting P/Giacobini--Zinner in 1985, and Giotto, Vega-1 and -2, Suisei and Sakigake examining P/Halley in 1986. Giotto went on to probe P/Grigg--Skjellerup in 1992. Although these missions provided valuable information about the structure of a cometary magnetosphere and its interaction with the solar wind at a given time, the fly-by nature of these missions did not enable the study of the evolution of this interaction as the heliocentric distance changes.
The Rosetta spacecraft spends most of its time close to the nucleus of comet 67P/Churyumov-Gerasimenko (67P/CG), scanning the cometary environment out to a maximum of 1500 kilometers only.
However, Rosetta has stayed in the vicinity of 67P/CG for two years giving us the unique opportunity to monitor and study {\it in situ} how the interaction evolves as the comet transforms from an almost atmosphereless object into an active nucleus \citep{glassmeier2007ssr}.   

 The Rosetta spacecraft completed its long voyage to comet 67P/CG in early August 2014 and recorded the first traces of cometary ions upon rendezvous. The first results from the plasma measurements made at 67P/CG describe the early phase of the comet's transformation. The first observations of cometary water ions and solar wind deflection were reported by \citet{nilsson2015science}, with the subsequent enhancement of comet water ion fluxes around comet 67P/CG down to a heliocentric distance of 2 au described in \citet{nilsson2015aa}. The local cold ion environment and its relation to the outgassing from the nucleus is described in \citet{goldstein2015grl} and \citet{edberg2015grl}. First reports on the magnetic field environment showed strong wave activity in the vicinity of the comet  \citep{richter2015ag}.

  At low activity, the solar wind is lightly mass-loaded with freshly ionized cometary particles. As these new{\bf-}born ions are accelerated by the solar wind electric field, energy and momentum are transferred from the solar wind to the coma. Considering the complete system ( i.e. the entire coma), the solar wind loses the energy gained by the cometary ions (often `referred to as `pick up' ions, as they are picked up by the solar wind).

\begin{figure}
   \begin{center}
      \includegraphics[width=.4\textwidth]{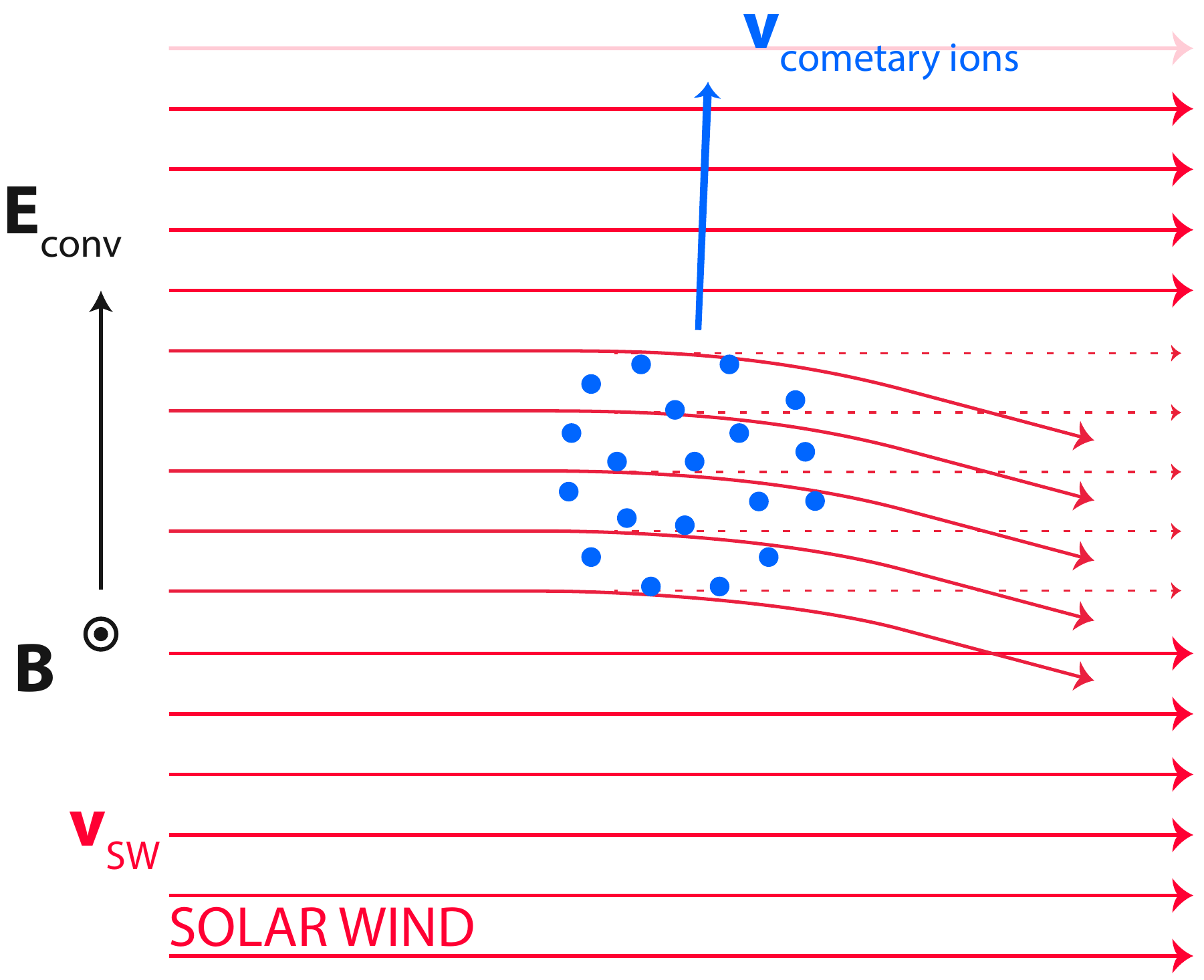}
      \caption{A simplistic view of light mass-loading: energy and momentum are transferred from the solar wind (red streamlines) to the cometary ions (blue dots) in a pick up process. Conservation of energy and momentum leads to a deceleration and deflection of the local solar wind. \label{lml}}
   \end{center}
\end{figure}

  The most basic expectations are illustrated in Figure \ref{lml}: the solar wind flow is deflected from the comet-sun line and slowed down, as new born cometary ions are accelerated. The gyroradius of the new born ions is much larger than the ion source, so these ions move essentially along the solar wind electric field within the ion source region. The newly created electrons on the other hand have a gyro-radius smaller than the source region, and can be expected to $\rm{E \times B}$ drift. This might lead to charge separation and, in turn, to new dynamical effects \citep{nilsson2015aa, behar2016grl}. \citet{coates2015jp} have discussed how such a situation near a low activity comet with a small coma may be more similar to barium release experiments than to higher activity comets. 

  The first observations at comet 67P/CG indeed showed solar wind deflection and water ion acceleration approximately orthogonal to the solar wind flow direction \citep{nilsson2015science}. The plasma dynamics of solar wind deflection at 67P/CG at low activity were further studied in \citet{broiles2015aa} and \citet{behar2016grl}. In these two studies, it was shown how the solar wind deflection direction and the direction of the acceleration of newborn cometary ions are both correlated with the local magnetic field direction, consistent with the transfer of momentum described here-above. 

  Once the comet activity has increased above the very low level of the initial observations, the cometary ion flow direction has a main anti-sunward component \citep{nilsson2015aa, behar2016grl}. \cite{behar2016grl} discuss this in terms of a polarisation electric field developing in the coma as electrons and ions respond differently to the solar wind electric field on scales below the ion gyroradius.

  We{\bf \sout{}} limit our study of the evolution of the interaction to the solar wind protons. This population reflects the interaction with the cometary ions experienced all along its trajectory through the coma. We expect the solar wind protons to exhibit both deflection and deceleration and we compare our expectations to the measured deflection and speed. In addition to the ion measurements, we consider the  local magnetic field, which is highlighted as one of the major drivers in the solar wind-comet interaction by \citet{broiles2015aa} and \citet{behar2016grl}.
  
  The {\it in situ} data recorded onboard Rosetta represent a single-point probe of the whole system, with little spatial coverage over time. Energy and momentum are transferred, via the electromagnetic field, over a larger volume. In order to get a large-scale view, we use a hybrid model of the coma-solar wind interaction. This model provides a global picture of the interaction, allowing us to get an estimate of the size of the region involved in energy and momentum transfer.

Considering the simplicity of our theoretical approach, we choose to define and use a fairly simple model as well, involving only the major components in the system. Cometary water ions are added to the solar wind flow through ionization of an isotropic neutral outflow. The model thus does not have a solid obstacle for the solar wind, nor does it have gravity or momentum transfer by collisions. By proceeding in this manner, we hope to isolate and identify the main drivers of the plasma dynamics at a comet, at low activity.

  We focus the study on the early period of the Rosetta mission, from the arrival at the nucleus (3.6 au) to a chosen distance of 2.2 au from the Sun. Closer to the Sun, a larger mass-loading of the solar wind results in a different dynamical regime in which, based on preliminary observations, either more radical or new effects occur. To mark this separation, we refer in this study to the phenomenon of light mass-loading, in opposition with a heavier mass-loading regime occurring later on.

%__________________________________________________________________

\section{Method}

\subsection{Instrument description}

   Particle data used in this work were produced by the Ion Composition Analyzer, part of the Rosetta Plasma Consortium (RPC-ICA). This instrument is an ion mass-energy spectrometer aimed at studying the interaction between the solar wind and positive cometary ions at comet 67P/CG \citep{nilsson2007ssr}. RPC-ICA data consist of count rates given in five dimensions, namely time, energy per charge, mass per charge, and incoming direction (two angles). Full energy scans are produced every 12 s and full angular scans are produced every 192 s. The energy spans from 10 eV up to 40 keV in 96 steps with a resolution $\frac{\delta E}{E} = 0.07$. The instrument field of view is $360 ^\circ \times 90 ^\circ$ ($azimuth\ \times\ elevation$), with a resolution of $22.5 ^\circ \times 5.0 ^\circ$. Mass is determined through a position detection system with 32 anodes, which we will refer to as mass channels. The radial position of ions on the detector plate is a function of both mass and energy.

   The magnetometer (RPC-MAG, \cite{glassmeier2007ssr-MAG}) measures the three components of the magnetic field vector, with a frequency of 20 vectors per second. The measurement range is $\pm$ 16384 nT with an accuracy of 31 pT. RPC-MAG is mounted on a 1.5 m long boom in order to minimise the impact of spacecraft-generated disturbance fields.
   
   In this work, the magnetic field amplitude was averaged over 10 hours of data.

\subsection{Particle data analysis}

\begin{figure}
   \begin{center}
      \includegraphics[width=.5\textwidth]{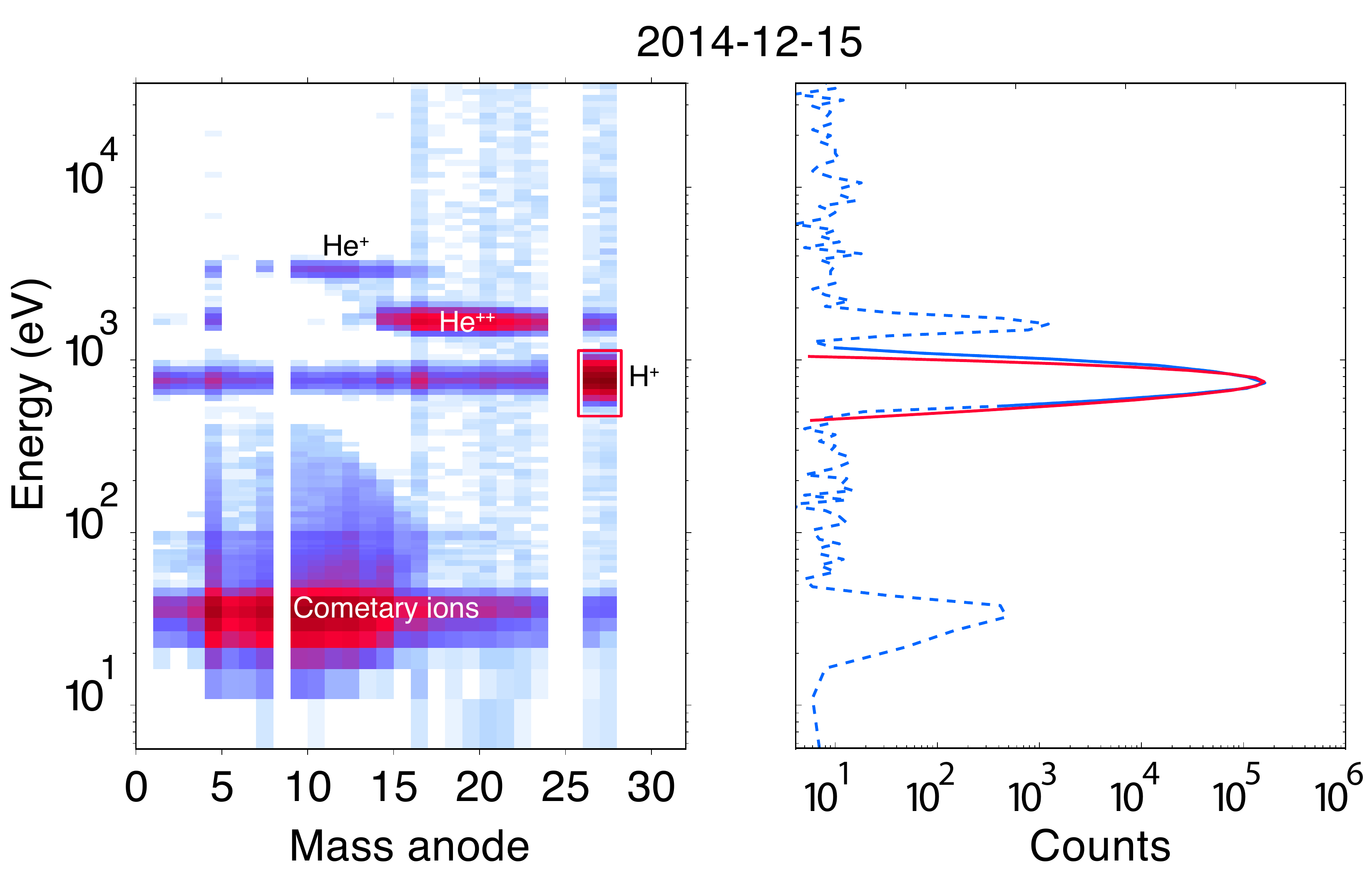}
      \caption{One example of an energy-mass matrix (left panel) and the corresponding proton fit (right panel). \label{EM}}
   \end{center}
\end{figure}

In order to characterise the protons detected by the instrument, the very first step was to identify and separate them from the rest of the observed ions. On the left panel of Figure \ref{EM}, counts integrated over one day are given as function of mass channel and energy, and different species are identified on this energy-mass matrix. The strongest signal was acquired for protons, and over 90\% of proton counts were detected in mass channels 26 and 27 (surrounded by dead mass channels). During the period covered by the study, protons were always separated from other species in the energy dimension, at the daily time scale. In the left panel of Fig.~\ref{EM}, protons (the lightest particles) form the rightmost population, at highest mass channels. At about twice the energy and shifted to the left $\rm{He^{2+}}$ particles are found, followed by $\rm{He^{+}}$ particles at about four times the proton energy and further to the left (lower mass channels). Cometary ions are found in the lower left corner, corresponding to lower energy and higher mass.

We have manually selected the proton energy, one selection per day. This selection was then used for all full angular scans (of 192 s) during each day.
The selection window which identifies proton counts and rules out the rest of the counts is defined for one day in energy-mass space, as follows: only mass channels 26 and 27 were considered, and we use an energy interval manually selected. This resulted in the red rectangle window over-plotted on the left panel (Fig. 2).

To determine the average direction of the proton flow, we computed full angular distributions (azimuth-elevation) with all counts in the selection window included. We calculated the direction of the bulk velocity in the body-Centered Solar EQuatorial (CSEQ) reference frame. The $x_{CSEQ}$ axis is along the sun-comet line, pointing to the Sun. We computed the solar wind deflection as the angle between $-x_{CSEQ}$ and the flow direction. The same method is presented with more details in \citet{behar2016grl}. 

This resulted in one series of deflection angles per day. The time resolution is 192s. Later on, we calculated the median of each series. In Section 3 we work with the time series of median deflection values, one median value per day.

On the right panel of Figure \ref{EM}, a normal distribution (solid red line) is fitted to the selected protons (solid blue line), along the energy dimension. In this example, the distribution is integrated over one day. To capture fast variations of the proton energy distribution, the fit parameters were calculated at the full angular scan time resolution (192s). We collected the central energy value, the height and the Full Width at Half Maximum (FWHM). This resulted in a time series of fit parameters, one series per day. The time resolution was 192s. Later on, we also calculate the median of each daily series, and work in Section 3 with a series of median fit parameters.

\subsection{Model}\label{sec:model}
To model the interaction between the comet 67P/CG and the solar wind plasma, we used a self-consistent hybrid plasma model where we included a production of cometary ions. In the hybrid approximation, ions are treated as particles, and electrons as a massless fluid. Below we present the governing equations for the solver and the comet model.  See \citet{Holmstrom2010, Holmstrom2013asp} for more information about the solver.

The trajectory of a particle, $\mathbf{r}\left(t\right)$ and $\mathbf{v}\left(t\right)$, with charge $q$ and mass $m$, is given by the solution of the equation of motion with the Lorentz force, $\mathbf{F}$:
\begin{equation}
 \frac{d\mathbf{r}}{dt} = \mathbf{v} , \; \; \; \; \frac{d\mathbf{v}}{dt} = \frac{\mathbf{F}}{m} =\frac{q}{m} \left( \mathbf{E}' + \mathbf{v} \times \mathbf{B} \right),
\end{equation}
with $\mathbf{E}' = \mathbf{E} - \eta \mathbf{J}$ to conserve momentum since electrons are massless \citep{Bagdonat2002jcp}, where $\mathbf{E} = \mathbf{E}\left(\mathbf{r},t\right)$ is the electric field, $\mathbf{B} = \mathbf{B}\left(\mathbf{r},t\right)$ is the magnetic field, and $\mathbf{J} = \mu_0^{-1} \nabla \times \mathbf{B}$ is the current density.

The electric field is not unknown, and is calculated by
\begin{equation}
 \mathbf{E} = \frac{1}{\rho_\mathrm{i}} \left( -\mathbf{J}_\mathrm{i} \times \mathbf{B} + \mathbf{J} \times \mathbf{B} - \nabla p_\mathrm{e} \right) + \eta \mathbf{J},
 \label{eq:ohm}
\end{equation}
where $\rho_\mathrm{i}$ is the ion charge density, $\mathbf{J}_\mathrm{i}$ is the ion current density, $p_\mathrm{e}$ is the electron pressure, and $\eta$ is the resistivity.

The gradient of the electron pressure is calculated by imposing quasi-neutrality and a polytropic index, $\gamma_\mathrm{e}$. In this study we chose an adiabatic index, corresponding to $\gamma_\mathrm{e}=5/3$.

Faraday's law is used to advance the magnetic field in time,
\begin{equation}
 \frac{\partial \mathbf{B}}{\partial t} = -\nabla \times \mathbf{E}.
 \label{eq:faraday}
\end{equation}

In vacuum regions, that are defined by the number density of ions being less than a given value, $n < n_\mathrm{min}$, we set $1/\rho_\mathrm{i} = 0$ in Eq.~\ref{eq:ohm}, and Faraday's law is reduced to solving the magnetic diffusion equation,
\begin{equation}
  \frac{\partial \mathbf{B}}{\partial t} = \frac{\eta}{\mu_0} \nabla^2 \mathbf{B}.
    \label{eq:induction}
\end{equation} 

A constraint on the time step has been inferred since the field cannot diffuse more than one cell per timestep,
 \begin{equation}
  \Delta t < \frac{ \mu_0 \left( \Delta x \right) ^2 }{ 2 \eta },
  \label{eq:cfl1}
 \end{equation} 
 where $\Delta t$ is the time step and $\Delta x$ is the cell size.
The time step is for moving the particles (ions).
The electromagnetic fields can be updated more frequently (subcycled) since it is usually computationally cheaper to update the fields compared to moving all the particles. 

For a comet, when neglecting gravity and assuming a constant outflow velocity, the total flux of non-collisional water vapor will be constant through any spherical shell around the nucleus at distance $r$. 
This is called the Haser model \citep{haser1957}, and is described below to a first order approximation that neglects neutral bi-products once created.

The number density of water vapor, $n$, as a function of the distance, $r$, from the comet nucleus is 
\begin{equation}
 n_{\mathrm{H}2\mathrm{O}} \left( r \right) = \frac{Q}{4 \pi r^2 \, u},
\end{equation}
where $Q$ is the production rate of water vapor, and $u$ is the mean velocity of water vapor in the radial direction.
However, if one accounts for losses (mainly due to photodissociation), the flux will decrease exponentially with time, $t= r/u$, as the molecules move outwards from the nucleus, and one gets
\begin{equation}
\label{eq:haser_nden}
 n_{\mathrm{H}2\mathrm{O}} \left( r \right) = \frac{Q}{4 \pi r^2 \, u} \, \exp \left( -\frac{\nu_\mathrm{d} \, r}{u} \right),
\end{equation}
where $\nu_\mathrm{d}$ is the photodestruction rate of water vapor.

The water vapor ionizes with a certain ionization rate, $\nu_\mathrm{i}$, and creates water ions, H$_2$O$^+$.
The water ion production rate as a function of distance becomes
\begin{equation}
\label{eq-ion-production-rate}
 q_\mathrm{i} \left( r \right) = \nu_\mathrm{i} \, n_{\mathrm{H}2\mathrm{O}} \left( r \right), %\; [\mathrm{m}^{-3} \, \mathrm{s}^{-1}],
\end{equation}
where, in the implementation, the number density of water, $n_{\mathrm{H}2\mathrm{O}}$, is taken at the center of each grid cell for each time step, generating the prescribed amount of ions at random positions in that cell.

Note that neglecting the neutral bi-products of water once they are created will barely change the density of neutral water if the mean-free-path due to photodissociation, $u/\nu_d$, is much larger than the size of the simulation domain, which is true for all cases studied in this paper. Other models used by, for example, \citet{hansen2007ssr,koenders2015pss}, have similar first order approximations that instead neglect the creation of neutral bi-products.

\subsubsection{Coordinate System and Simulation Box}

The coordinate system we use in the hybrid model is a body-centered coordinate system.
It is centered in the middle of 67P/CG.

The $x$-axis is pointing towards the Sun, with solar wind flowing in the $-x$ direction, making it the same $x$ coordinate as in the CSEQ reference frame, which is the only coordinate addressed in the observations.
The $z$-axis is pointing in the direction of the ambient convective electric field,
and the $y$-axis completes the right-handed system.
We assume that the IMF has a Parker spiral configuration, that is the IMF lies in the $xy$-plane.

The convective electric field is given by $\textbf{E}_0 = -\textbf{v}_0 \times \textbf{B}_0$, where $\textbf{v}_0$ is the undisturbed solar wind bulk velocity, and $\textbf{B}_0$ is the IMF which is initially homogeneous everywhere.

The simulation domain is given by $|x| < 6\cdot 10^3$ km, $|y| < 12\cdot 10^3$ km, $|z| < 18\cdot 10^3$ km, with a cellsize  $\Delta x = 125$ km.

\subsubsection{Model and simulation parameters}\label{sec:parameters}

We set the plasma resistivity to $\eta_\mathrm{p} = 1.6 \cdot 10^4$ Ohm~m, to dampen numerical oscillations.
We assume a vacuum resistivity of $\eta_\mathrm{v} = 2.5 \cdot 10^5$ Ohm~m, which is used when solving the diffusion equation of Faraday's law (Eq.~\ref{eq:induction}) for regions of a number density less than the arbitrarily chosen $n_\mathrm{min} = n_0 / 16$, where $n_0$ is the ambient solar wind proton number density.
To summarize:
\begin{equation}
 \eta = 
\begin{cases}
\eta_\mathrm{v}, &\; \mathrm{for} \, n<n_\mathrm{min},
\\
\eta_\mathrm{p}, &\; \mathrm{otherwise}.
\end{cases}
\end{equation}

The electromagnetic fields were updated 20 times for each time step, $\Delta t = 3.5 \cdot 10^{-2}$. The simulations were run for a total of 120~s to reach steady-state.

The water production rate of a comet changes with distance from the sun. We used $Q = 1.14 \cdot 10^{29} \cdot R^{-7.06}$ [s$^{-1}$], where $R$ is the distance to the sun in au \citep{simon2016aa}.
The neutral expansion velocity is observed to be relatively constant around $u=$ 0.7 km/s \citep{simon2016aa}, so we used that value.

Water ions are produced in the simulation according to Eq.~\ref{eq-ion-production-rate}.
We used scaled values, $1/R^2$ with heliocentric distances, for photoionization and photodestruction from \citet{crovisier1989aa}, where a mean value was taken between active and quiet Sun conditions.
At 1 au this gives a photoionization rate of $\nu_\mathrm{i} = 6.5 \cdot 10^{-7}$ s$^{-1}$ and a photodestruction rate of $\nu_\mathrm{d} = 1.78 \cdot 10^{-5}$ s$^{-1}$ \citep{crovisier1989aa}.

In the model we used typical solar wind conditions at 1 au from \citet{cravens2004} scaled to the heliocentric distances of 67P/CG.
The undisturbed solar wind plasma parameters at 1 au are a bulk velocity of $v_{0}$ = 400~km/s in $-\hat{x}$ (neglecting aberration), a number density of protons being $n_0$ = 5~cm$^{-3}$, and a temperature of ions and electrons being constant at $T_\mathrm{i} = 0.5 \cdot 10^5$~K and $T_\mathrm{e} = 1 \cdot 10^5$~K, respectively.
The interplanetary magnetic field (IMF) has a magnitude of 7 nT with a Parker spiral angle of $\chi = 45^\circ$.

To model the evolution of the solar wind interaction with the comet as the comet approaches the Sun, we chose six different heliocentric distances ranging from when Rosetta arrived at the comet at around $R = 3.6$ au, to $R = 2.0$ au.
Six cases are summarized in Tables \ref{tab:parameters1} and \ref{tab:parameters2}, that can be compared with similar cases modeled by \citet{hansen2007ssr}.

\begin{table}[ht]
\caption{Solar wind conditions used in the model.}\label{tab:parameters1}
\centering
\begin{tabular}{l l l l l l}
\hline
 Case & R[au] & $n_0$[cm$^{-3}$] & $v_0$[km/s] & $B_0$[nT] & $\chi$[$^\circ$]  \\
\hline
 1 & 3.6 & 0.385 & 400 & 1.4 & 74 \\
 2 & 3.25 & 0.45 & 400 & 1.6 & 73 \\
 3 & 3.05 & 0.55 & 400 & 1.7 & 72  \\
 4 & 2.7 & 0.7 & 400 & 2.0 & 70  \\
 5 & 2.35 & 0.9 & 400 & 2.3 & 67  \\
 6 & 2.0 & 1.25 & 400 & 2.8 & 63  \\
\hline
\end{tabular}
\end{table}

\begin{table}[ht]
\caption{Cometary parameters used in the model.}\label{tab:parameters2}
\centering
\begin{tabular}{l l l l l l}
\hline
 Case & R[au] & $u$[km/s] & $Q$[s$^{-1}$] & $\nu_\mathrm{i}$[s$^{-1}$] & $\nu_\mathrm{d}$[s$^{-1}$] \\
\hline
 1 & 3.6 & 0.7 & $1.3 \cdot 10^{25}$ & $5.0 \cdot 10^{-8}$ & $1.4 \cdot 10^{-6}$ \\
 2 & 3.25 & 0.7 & $2.8 \cdot 10^{25}$ & $6.2 \cdot 10^{-8}$ & $1.7 \cdot 10^{-6}$ \\
 3 & 3.05 & 0.7 & $4.3 \cdot 10^{25}$ & $7.0 \cdot 10^{-8}$ & $1.9 \cdot 10^{-6}$ \\
 4 & 2.7 & 0.7 & $1.0 \cdot 10^{26}$ & $8.9 \cdot 10^{-8}$ & $2.4 \cdot 10^{-6}$ \\
 5 & 2.35 & 0.7 & $2.7 \cdot 10^{26}$ & $1.2 \cdot 10^{-7}$ & $3.2 \cdot 10^{-6}$ \\
 6 & 2.0 & 0.7 & $8.5 \cdot 10^{26}$ & $1.6 \cdot 10^{-7}$ & $4.5 \cdot 10^{-6}$ \\
\hline
\end{tabular}
\end{table}

\section{Results}
 
\begin{figure*}
   \begin{center}
      \includegraphics[width=\textwidth]{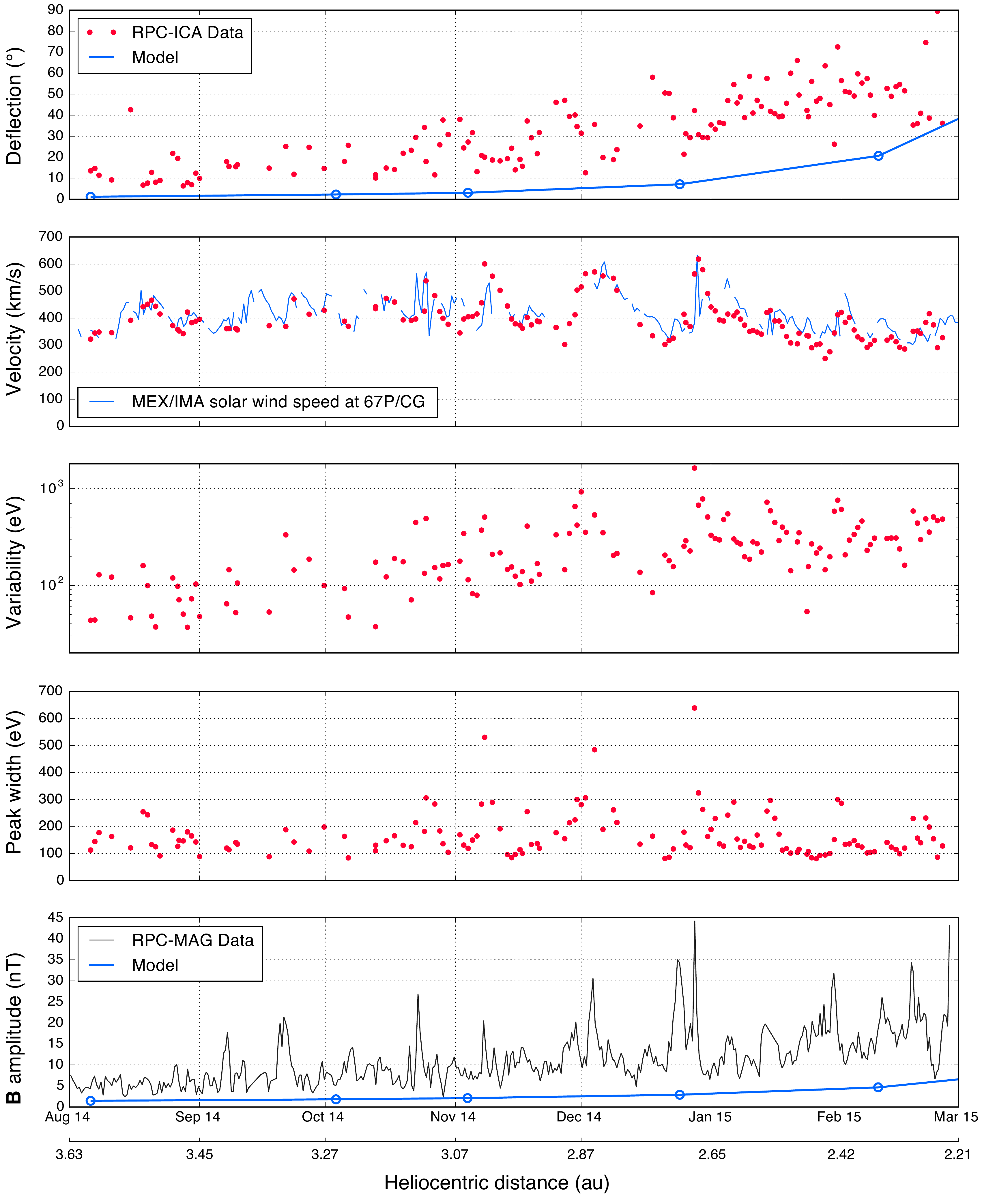}
      \caption{Time series for: deflection angle (first panel), speed (second panel), energy distribution width (third panel), peak centre variability (fourth panel) and magnetic field amplitude (last panel). RPC-ICA data are given by red dots - every dot is a daily median. Peak width and peak central value are given by the fit parameters. RPC-MAG magnetic field amplitude is averaged over 10 hours. Blue lines are either results from the hybrid model (first and last panels) or results from the MEX solar wind speed propagation (second panel from top) \label{results}}
   \end{center}
\end{figure*}

 The evolution of the angle between the proton flow direction and the comet-sun line is given in Figure \ref{results} (top panel) for both the model and the observations. We assume the upstream solar wind is flowing radially from the Sun, and the presented proton deflection angle is then the deflection relative to this assumed initial radial flow. 
 
 \citet{broiles2015aa} and \citet{behar2016grl} previously reported a significant deflection of the solar wind for precise cases, and we now consider a much larger time scale. The observed deflection increases smoothly with decreasing heliocentric distances, reaching a median daily value in the range of 30 to 90$^\circ$. The model displays the same trend, with a non-linear increase reaching a value of 40$^\circ$.

 One would expect the first observed deflection values in the time series to be very close to 0$^\circ$, since the mass-loading at the time is extremely light. This is seen in the model results, but the deflection is larger in the observations. A closer look at the observed data reveals that the Sun, as seen in the instrument field of view, was just a few degrees away from the spacecraft body. This partial obstruction of the solar wind flow and the way the instrument software handles it  (onboard computation) should have a significant influence on the computed deflection, until around mid-September 2014.
 
 By comparing panels 1 and 2 of Figure \ref{results}, one can see that the variability in deflection is anti-correlated with the measured proton velocity.\\
 
   To estimate the upstream solar wind speed at 67P/CG, we used solar wind speed measured by the Ion Mass Analyzer (IMA), which is one of the sensors in the ASPERA-3 instrument package \citep{barabash2006ssr} onboard Mars Express. IMA is an ion spectrometer almost identical to RPC-ICA. We propagated the speed from Mars to 67P/CG, with the assumption that solar wind conditions are identical along a Parker spiral, and remain unchanged over time. We calculate the delay for the Parker spiral to get from one body to the other. Thus delays can be positive as well as negative.
   
\begin{figure}
   \begin{center}
      \includegraphics[width=.5\textwidth]{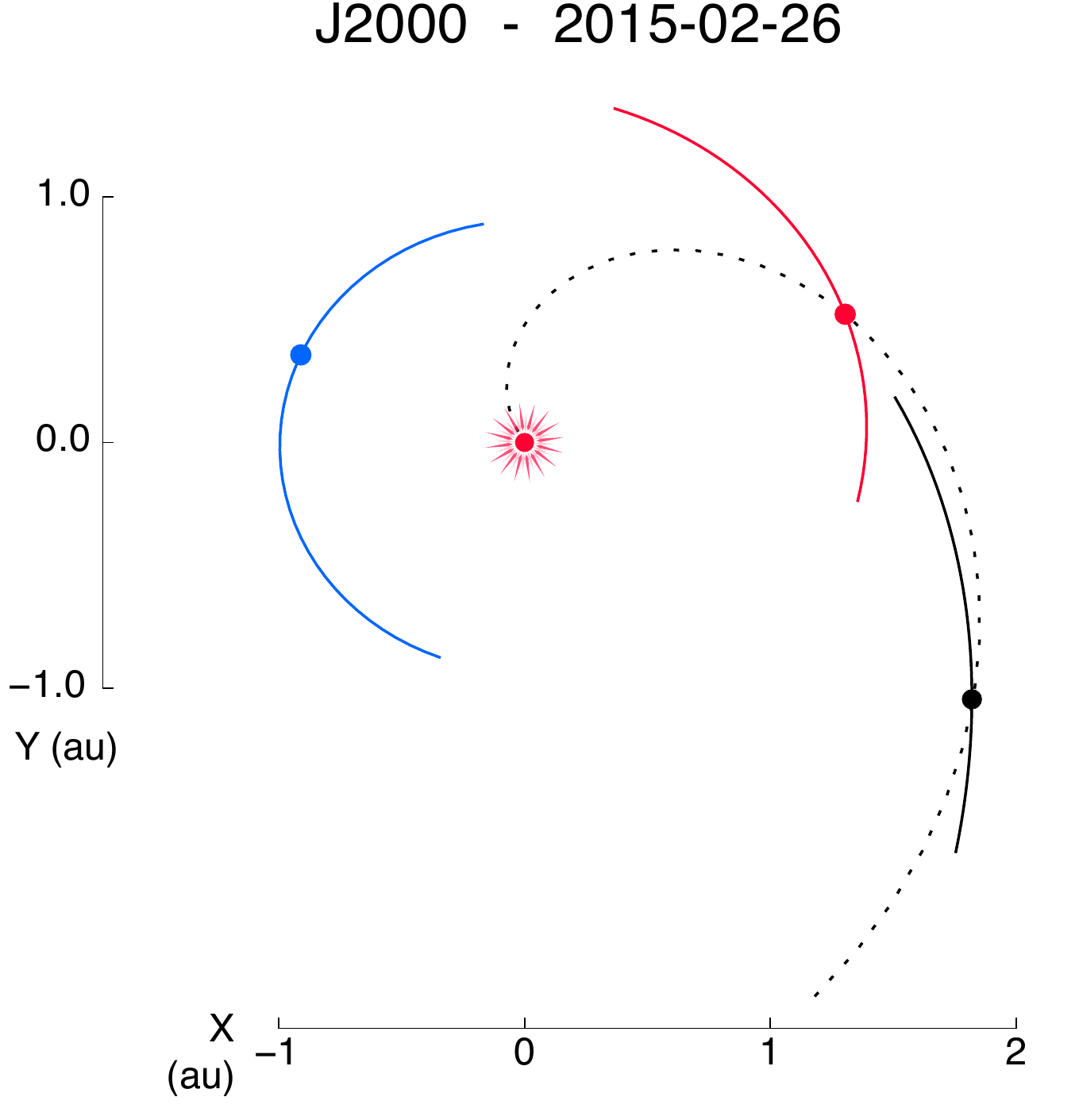}
      \caption{The position of Earth (blue), Mars (red), and 67P/CG (black) in J2000 coordinate system during 2015-02-26, with part of their orbits. The Parker spiral intersecting Mars for that day is also intersecting 67P/CG (dotted). \label{orb}}
   \end{center}
\end{figure}

   Over the period of the study, the propagation delay started at a high value of 12 days, meaning that what is measured at Mars at one time will be a good estimation at 67P/CG 12 days later. This delay becomes shorter than 4 days after December, to end up at values around 0: the Parker spiral intersects both bodies simultaneously, as shown in Figure \ref{orb}. In other words, this propagated speed should be more reliable when getting closer to the end of the period.

   The propagated solar wind speed is shown as a blue line in the second panel from the top in Figure \ref{results}. Propagating large variations in the speed results in large time periods not being covered ( gaps in the blue line): the propagation delay is a function of the solar wind speed, and two different events measured at Mars at two different times can arrive at the same time at the comet, or even in the opposite order of occurrence.

   We get a very good agreement between the estimated upstream speed and the speed measured inside the coma. RPC-ICA data time coverage gets better with time during this period. Thus after January, we have better statistics, and more reliable solar wind upstream speed (as pointed out above). The average deceleration after January is about -40km/s from estimated upstream speed to the point of measurement, but no clear trend can be discerned. The uncertainty on this deceleration estimation is rather large, extremely difficult to quantify, and does not allow us to judge whether or not deceleration is systematically observed during the later period. Some features measured at Mars and propagated to the comet are in fact not observed at the comet. Even though the expected delay taken into account when propagating the observations from Mars to the comet is small during the second half of the time period, the absolute distance between the two bodies is never less than 1.7 au . We make the assumption that solar wind conditions are identical along a Parker spiral, but it is obvious that the larger the distance between the two bodies, the worse this assumption gets.
   
\begin{figure}
   \begin{center}
      \includegraphics[width=.5\textwidth]{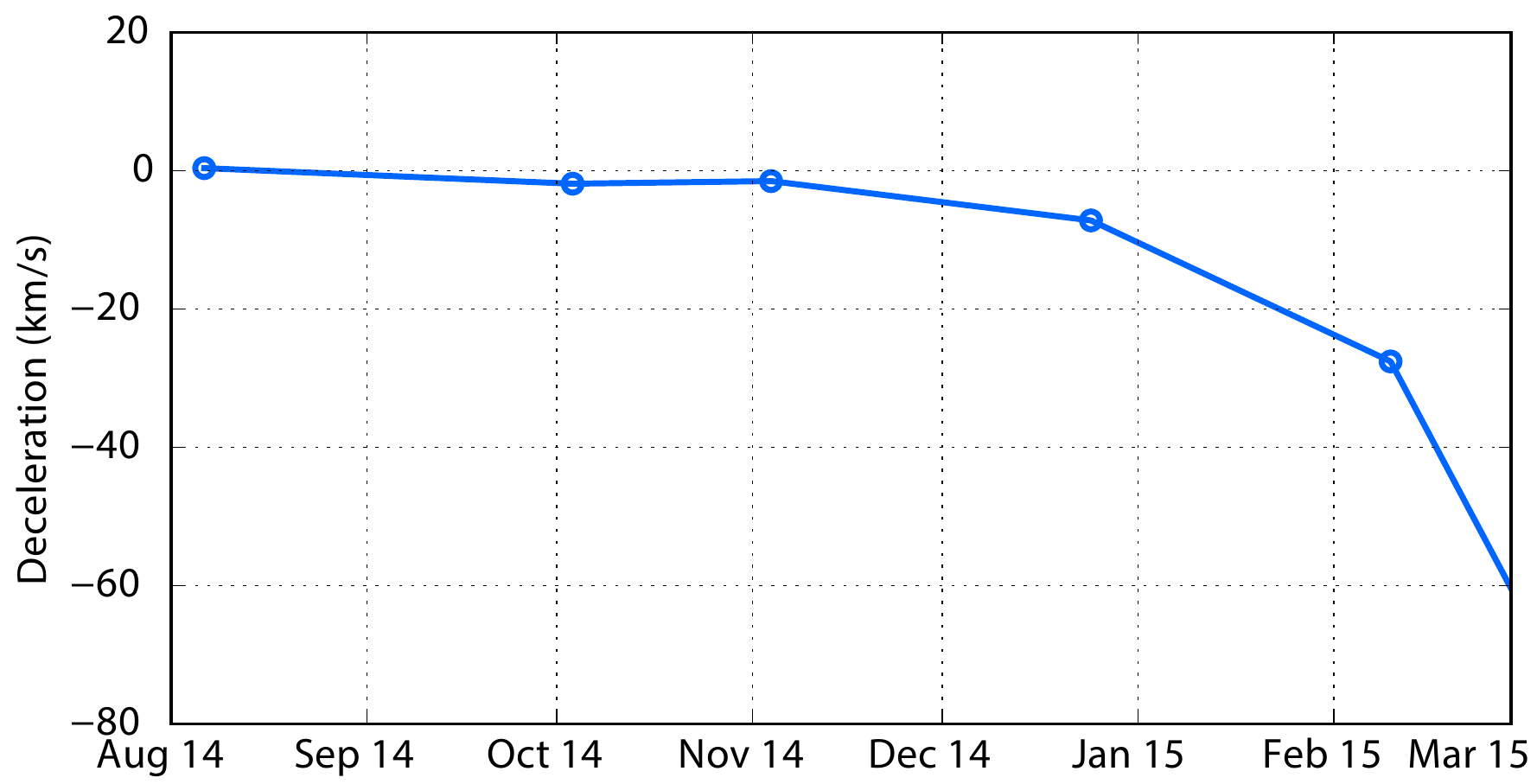}
      \caption{Deceleration seen in the model between a point upstream and the nucleus. The maximum deceleration is 60 km/s at 2.2 au. \label{decel}}
   \end{center}
\end{figure}

   The hybrid model shows a constant deceleration over the period, with a maximum deceleration of 60 km/s at 2.2 au (cf. Figure \ref{decel}). This is consistent with the range estimated from the data for the deceleration.\\

   It appears that with decreasing heliocentric distances, the variability of the proton speed (or energy) at the 192 s resolution increases, as seen in the 4th panel of Figure \ref{results}. The variability is given as the difference between the 5th and 95th centiles of the proton peak central values (in energy) observed during one day. At the beginning of our period of study, this variability is about 50 to 100 eV  (100 to 140 km/s). It reaches values of several hundreds to a thousand electron-volts at the end of this period. By comparing panels 2 and 4 of Figure \ref{results} one can see that the variability of the proton speed in the coma correlates with the upstream proton speed.\\

The FWHM of the fitted proton spectra, (Figure \ref{results}, 3rd panel) doesn't appear to correlate with heliocentric distance, but correlates with the upstream proton velocity.\\
   
   In the last panel of Figure \ref{results}, the magnetic field amplitude increases with time and decreasing heliocentric distance, from 5 nT $\pm$ 2nT to around 20 nT. The peaks in the magnetic field amplitude correlate well with peaks in the proton speed. The model (solid blue line in the same panel) also results in a similar relative increase, from 1.5 to 6.6 nT, though the magnetic field magnitude is lower. The model shows a less dramatic increase compared to the observations.\\

\begin{figure*}
   \begin{center}
      \includegraphics[width=.6\textwidth]{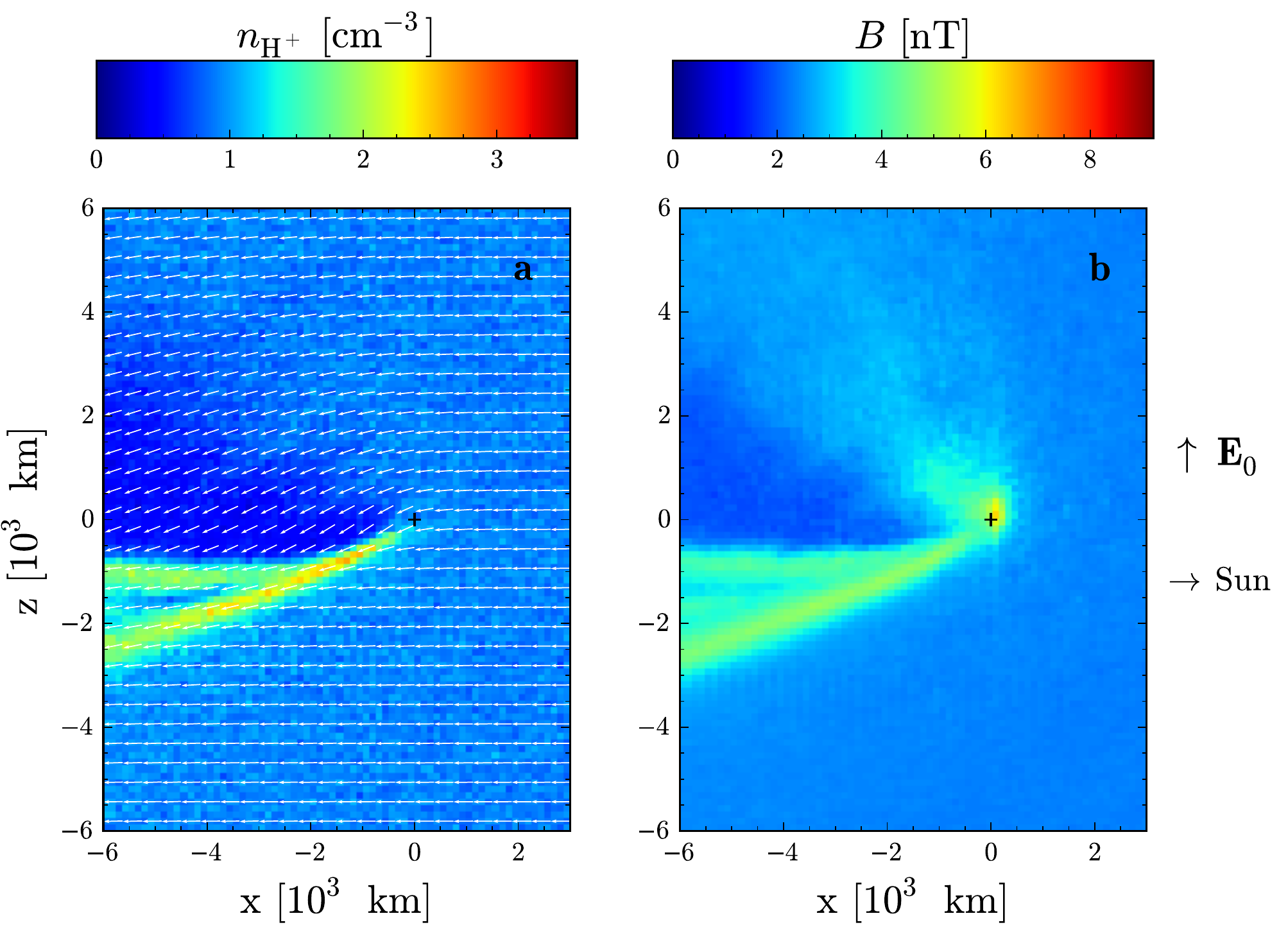}
      \caption{The number density of solar wind protons together with their bulk velocity as a vector field (\textbf{a}), and the magnetic field magnitude (\textbf{b}). Cuts through $y=0$ for the comet at 2.35 au, case five. The ambient solar wind number density is 0.9 cm$^{-3}$, and the ambient magnetic field is 2.3 nT with a Parker spiral angle of 67$^\circ$. The location of the comet is marked by black cross-hairs.} \label{modelresults-nB}
   \end{center}
\end{figure*}

Finally, the morphology of the interaction when the comet was at 2.35 au can be seen in the hybrid model results in Figure \ref{modelresults-nB}.

%_________________________________________________________________

\section{Discussion}

The solar wind observed inside the coma, close to the nucleus, is a good tracer of the interaction between the solar wind and the coma. The spacecraft altitude did not change significantly during the investigated time period; the spacecraft spent 90\% of the time below 100 km in altitude. Thus the variations we observe are to a large extent temporal, and not spatial, and the evolution of the proton flow parameters directly reflects the time evolution of the plasma environment. We expect the solar wind to be deflected and slowed down. Both the deflection and the deceleration should increase with increasing comet activity, or a decreasing heliocentric distance. Significant and increasing deflection is  observed during the whole period of study, and data from the later part is consistent with deceleration, but we cannot observe a clear trend in this deceleration.

\subsection*{Coma evolution}
All of the aspects of the protons we have been studying over this period present detectable changes that can be correlated with either the heliocentric distance or the upstream solar wind velocity. The decreasing heliocentric distance gives the main trend in the evolution of the deflection, the magnetic field amplitude, the energy variability and, in the model, the deceleration. The variability around this main trend seems to be associated with the upstream proton velocity: the proton peak width together with the proton energy variability present the best correlation with the upstream speed, but an anti-correlation with deflection is also seen for intermediate heliocentric distances. The main peaks in the measured magnetic field amplitude are also aligned with peaks in upstream velocity.

With the resolution we get after data analysis, no physical aspect other than the heliocentric distance and the upstream solar wind speed is needed to describe the evolution of the deflection and the energy distribution of the solar wind protons.

\subsection*{A near-orthogonal force on the protons}
Even though we noted that the degree of deceleration of the protons was difficult to assess from our data, we can clearly state that the protons are largely deflected, but not significantly slowed down. Thus the force applied on the observed protons must be near-orthogonal to their velocity, all along their trajectory to the instrument. The observed protons have not lost a significant amount of energy.

\subsection*{An unbalanced Lorentz force}
During the investigated time period from August 2014 to March 2015, the magnetic field amplitude $|{\bf B}|$ measured at the spacecraft increased to average values of approximately 20 nT. We therefore have an augmentation of $|{\bf B}|$ following the comet activity escalation, but also an augmentation of $|{\bf B}|$ along a proton trajectory, from upstream of the coma to the measurement point. In the undisturbed solar wind, the magnetic force $q{\bf v}_{\mathrm{H^+}} \times {\bf B}$ and the electric force $q{\bf E}$ are cancelling each other, the Lorentz force ${\bf F} = q({\bf E} + {\bf v}_{\mathrm{H^+}} \times {\bf B})$  is balanced and null.

It is of great interest to put this increase of $|{\bf B}|$ in opposition with the rather small deceleration of the solar wind. Without an opportune new configuration of the local electric field in both amplitude and direction, the Lorentz force applied on the protons is not equal to zero anymore. The solar wind protons are perturbed by this increase in $|{\bf B}|$ along their trajectory. A significant part of their motion can now be seen as gyrating rather than a true bulk drift. The fact that the magnetic field is enhanced without a corresponding deceleration of the protons indicates that the protons are no longer coupled to the magnetic field, or a significant part of their energy is now in a gyromotion, with a reduced bulk drift.

\subsection*{Transfer of momentum and energy - the larger picture}

 We observe a force applied on the solar wind protons in the coma, a force mainly orthogonal to their bulk velocity. This force is therefore efficient in changing their momentum, but acts less efficiently on their energy, that is, their speed. Both momentum and energy are transferred to the water ions through electromagnetic fields. \citet{haerendel1982zna} and \citet{brenning1991jgr} for example, in the context of barium releases, depict a momentum transfer along Alfv\'{e}n waves propagating from the cloud (artificial coma).
 
 So the previous observation raises the question about the actual shape of the regions where energy and momentum exchange takes place. In particular, these regions are not necessarily identical.
 
 Since we only have a one-point measurement to assess the situation, we must turn to the modeled interaction to see the larger picture.\\

In the hybrid model, we can separate the Lorentz force, $\mathbf{F} = q(\mathbf{E}' + \mathbf{v}_{\mathrm{H}^+} \times \mathbf{B})$ into two components: the force acting parallel to the solar wind proton bulk flow, $\mathbf{F}_\parallel = \hat{\mathbf{v}}_{\mathrm{H}^+} \left( \mathbf{F} \cdot \hat{\mathbf{v}}_{\mathrm{H}^+} \right)$, and the force acting perpendicular to the flow, $\mathbf{F}_\perp = \mathbf{F} - \mathbf{F}_\parallel$. The perpendicular component acts to make the bulk of the solar wind deflect, while the parallel component acts to change the solar wind speed.

\begin{figure*}
   \begin{center}
      \includegraphics[width=.6\textwidth]{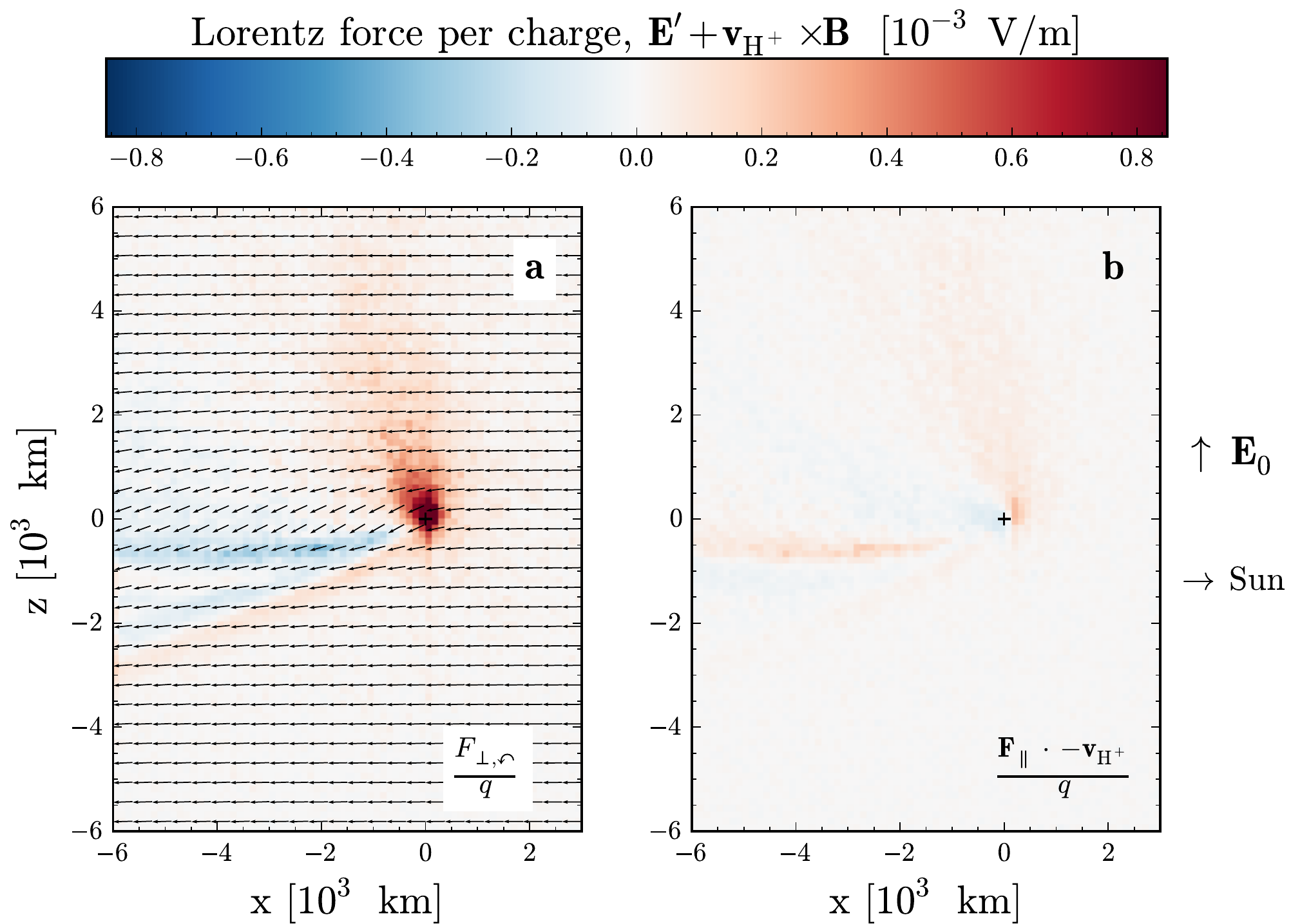}
      \caption{The Lorentz force per charge acting on the bulk of the solar wind protons of a cut through $y=0$ for the comet at 2.35 au, case 5. The solar wind proton bulk velocity is given as a vector field (\textbf{a}). The force acting perpendicular to the solar wind bulk velocity, $\mathbf{v}_{\mathrm{H^+}}$, deflects the solar wind protons (\textbf{a}), where a positive value is taken as counter-clockwise gyration by convention. The force parallel to the solar wind bulk velocity changes the solar wind speed (\textbf{b}), where a positive value results in a deceleration. The ambient convective electric field is 0.85 V/km. The location of the comet nucleus is marked by black cross-hairs.} \label{modelresults}
   \end{center}
\end{figure*}

Hybrid model results for the comet located at 2.35 au, case 5, show that the Lorentz force, $\mathbf{F}$, is primarily making the bulk of the solar wind protons gyrate and thus deflect towards the direction opposite of the convective electric field ($-\hat{z}$).
The perpendicular force, $\mathbf{F}_\perp$, deflects the protons in a counter-clockwise manner in the $xz$ plane, seen in Figure~\ref{modelresults}a, where the bulk velocity of the solar wind protons are given as a vector field.

The parallel force, $\mathbf{F}_\parallel$, which is slowing the solar wind down, is much weaker (see Figure~\ref{modelresults}b, where a positive value corresponds to deceleration). This is in agreement with the in-situ observations.\\

The regions of deflection (perpendicular force) and deceleration (parallel force) are of comparable sizes and shape, but with different strengths. It is interesting to note that the gyroradius of a pick up ion in the undisturbed solar wind is $3 \cdot 10^5$ km, which is much larger than the interaction region of about $10^3$ km.
The interaction region leading to deflection and deceleration of the solar wind protons corresponds to the region where the newly ionized water is accelerating along the convective electric field, $\mathbf{E}_0$.\\

%______________________________________________________________

\section{Conclusion}

  As the heliocentric distance decreases from 3.6 to 2.2 au, the observed solar wind is increasingly deflected, up to 90$^\circ$. The modeled interaction results in the same evolution of the deflection angle, with a lower maximum value than observed.
   
   In contrast with this strong deflection, the observed solar wind is not significantly slowed down, with an estimated deceleration of 40 km/s at heliocentric distances between 2.65 and 2.2 au. The modeled deceleration is consistent with the observations. \\
   
    The strong proton deflection is the most obvious signature of mass-loading at a small comet, while little deceleration is observed. This may also have important implications for other objects where the interaction region is small compared to the gyro radii of pick up ions. The interplanetary magnetic field at the orbit of Pluto is very small and the pick up ion gyro radius correspondingly large, in the order of 1 million km. If there is significant mass-loading of the solar wind due to an extended atmosphere upstream of the bow shock at Pluto, this interaction is likely similar to that of a small scale comet. The New Horizons spacecraft observed very little deceleration of the solar wind at its flyby of Pluto \citep{bagenal2016science, mccomas2016jgr}, while no clear data on solar wind deflection has been published. \citet{mccomas2016jgr} reported a clear pressure-related plasma boundary forming between the ionosphere and the solar wind at Pluto, so in this respect Pluto behaves similarly to other unmagnetised planets such as Mars and Venus. \\

      As contrast to the single-point limitation of the measurements, the model enables us to describe the complete picture of the interaction. In the model, the region where energy is transferred and the region where momentum is transferred are close to identical. The force acting on the solar wind protons has a main component orthogonal to their bulk velocity. This  confirms and completes the picture we get from the observations, in which solar wind protons are seen as almost gyrating.

   The region where momentum and energy are transferred from the solar wind to the coma is centered on the nucleus, with a dimension in the order of $10^3$ km. But this localized interaction has significant effects on the downstream solar wind flow over much larger spatial scales. The solar wind is in fact seen piling up in the (-z, -x) quadrant, a region where neither momentum nor energy are significantly transferred. This happens where the deflected solar wind intersects the nearly undisturbed solar wind.

\begin{acknowledgements}

This research was conducted using resources provided by the Swedish National Infrastructure for Computing (SNIC) at the 
High Performance Computing Center North (HPC2N), Ume\aa\ University, Sweden.

The software used in this work was in part developed by the DOE NNSA-ASC OASCR Flash Center at the University of Chicago.

The hybrid solver is part of the openly available FLASH code and can be downloaded from http://flash.uchicago.edu/.

The simulation results are available from the corresponding author on request.

This work was supported  by the Swedish National Space Board (SNSB) through grants 108/12, 112/13, 96/15 and 94/11. 

We acknowledge the staff of CDDP and IC for the use of AMDA and the RPC Quicklook database (provided by a collaboration between the Centre de Donn\'{e}es de la Physique des Plasmas (CDPP) supported by CNRS, CNES, Observatoire de Paris and Universit\'{e} Paul Sabatier, Toulouse and Imperial College London, supported by the UK Science and Technology Facilities Council). We are indebted to the whole Rosetta mission team, Science Ground Segment and Rosetta Mission Operation Control for their hard work making this mission possible. 
\end{acknowledgements}


\bibliography{cometLib}
\bibliographystyle{aa}

\end{document}